\documentclass[10pt,letterpaper,twocolumn]{article} %% two column, final layout
\usepackage{ol2}
\usepackage[draft]{hyperref}
\usepackage{amsmath}
\usepackage{graphicx}
\usepackage{floatrow}
\graphicspath{{pict/}{./}}
\usepackage[font={small}]{caption}

\usepackage{bm}%

\newcommand\rpict[1]{\ref{#1}}

\newcounter{Fig}

\newcommand{\be}{\begin{equation}}
\newcommand{\ee}{\end{equation}}

\begin{document}
\twocolumn[
\title{Invisible nanowires with interferencing electric and toroidal dipoles}
\author{Wei Liu,$^{1,*}$ Jianfa Zhang,$^{1}$ Bing Lei,$^{1}$ Haojun Hu,$^{1}$ and Andrey E. Miroshnichenko$^2$}
\address{
$^1$College of Optoelectronic Science and Engineering, National University of Defense
Technology, Changsha, China\\
$^2$Nonlinear Physics Centre, Australian National University,
Acton, ACT 0200, Australia\\
$^*$Corresponding author: wei.liu.pku@gmail.com
}
%--------------------------------------------------------------------
\begin{abstract}
By studying the scattering of  normally incident plane waves by a single nanowire, we
reveal the  indispensable role  of toroidal multipole excitation in multipole expansions of radiating sources. It is found that for both \textit{p}-polarized and \textit{s}-polarized incident waves, toroidal dipoles can be effectively excited within homogenous dielectric nanowires in the optical spectrum regime.  We further demonstrate that the plasmonic core-shell nanowires can be rendered invisible through destructive interference of the electric and toroidal dipoles, which may inspire many nanowire based light-matter interaction studies, and incubate biological and medical applications that require non-invasive detections and measurements.
\end{abstract} \ocis{290.5839 % Scattering, invisibility
240.6680,   %Surface plasmons,
260.5740.   %Resonance
%350.5500,   %Propagation
} ] %% activate for two-column option
%\maketitle
%---------------------------------------------------------------------
In conventional Cartesian multipole expansion method, the radiation sources have been decomposed into electric and magnetic multipoles, where the contributions of toroidal components are usually overlooked~\cite{jackson1962classical}.  In contrast to electric and magnetic counterparts, the toroidal multipole (TM) can be viewed as poloidal current configurations, which flow on the surface of a torus along its meridians, or as a series of magnetic dipoles (enclosed circulating currents) aligned along an enclosed path, which interacts only with the time derivatives of the incident fields~\cite{Dubovik1990_PR,Radescu2002_PRE}.  Though it is long known that the conventional Cartesian multipole expansion method is incomplete as the indispensable role played by TMs is not discussed~\cite{Dubovik1990_PR,Radescu2002_PRE}, TMs have only attracted the significant attention since the recent experimental demonstration of the toroidal responses of metamaterials consisting of specially arranged split ring resonators~\cite{Kaelberer2010_Science}. Following this work, shortly afterwards in various other structures and spectrum regimes the excitation of TMs have also been identified, which are expected to be of fundamental importance for various applications including biosensing, nanoantennas, photovoltaic devices and so on~\cite{dong2012toroidal,Huang2012_OE,Ogut2012_NL,Fedotov2013_SR,Fan2013_lowloss,dong2013all,savinov2014toroidal,Basharin2014_arXiv}.
%-------------------------------------------------------------------------------
\begin{figure}
\centerline{\includegraphics[width=7cm]{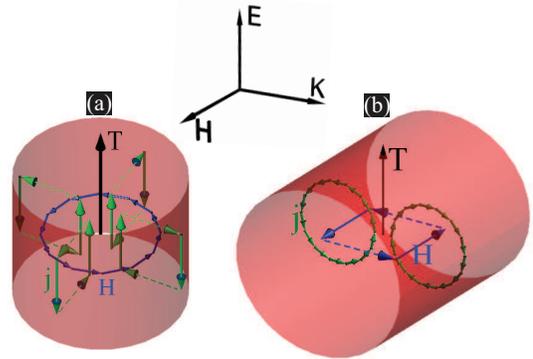}}\caption{(\small (Color online) Schematic of the nanowire scatterings of norammly incident (a) $s$-polarized and (b)  $p$-polarized plane waves. The TMs excited within the nanowires are illustrated through the current \textbf{j} [or electric field $\textbf{E}$ according to Eq.(\ref{current})] and magnetic field $\textbf{H}$ distributions.}
\label{fig1}
\end{figure}
%-------------------------------------------------------------------------------

%-------------------------------------------------------------------------------
\begin{figure*}
\centerline{\includegraphics[width=15.5cm]{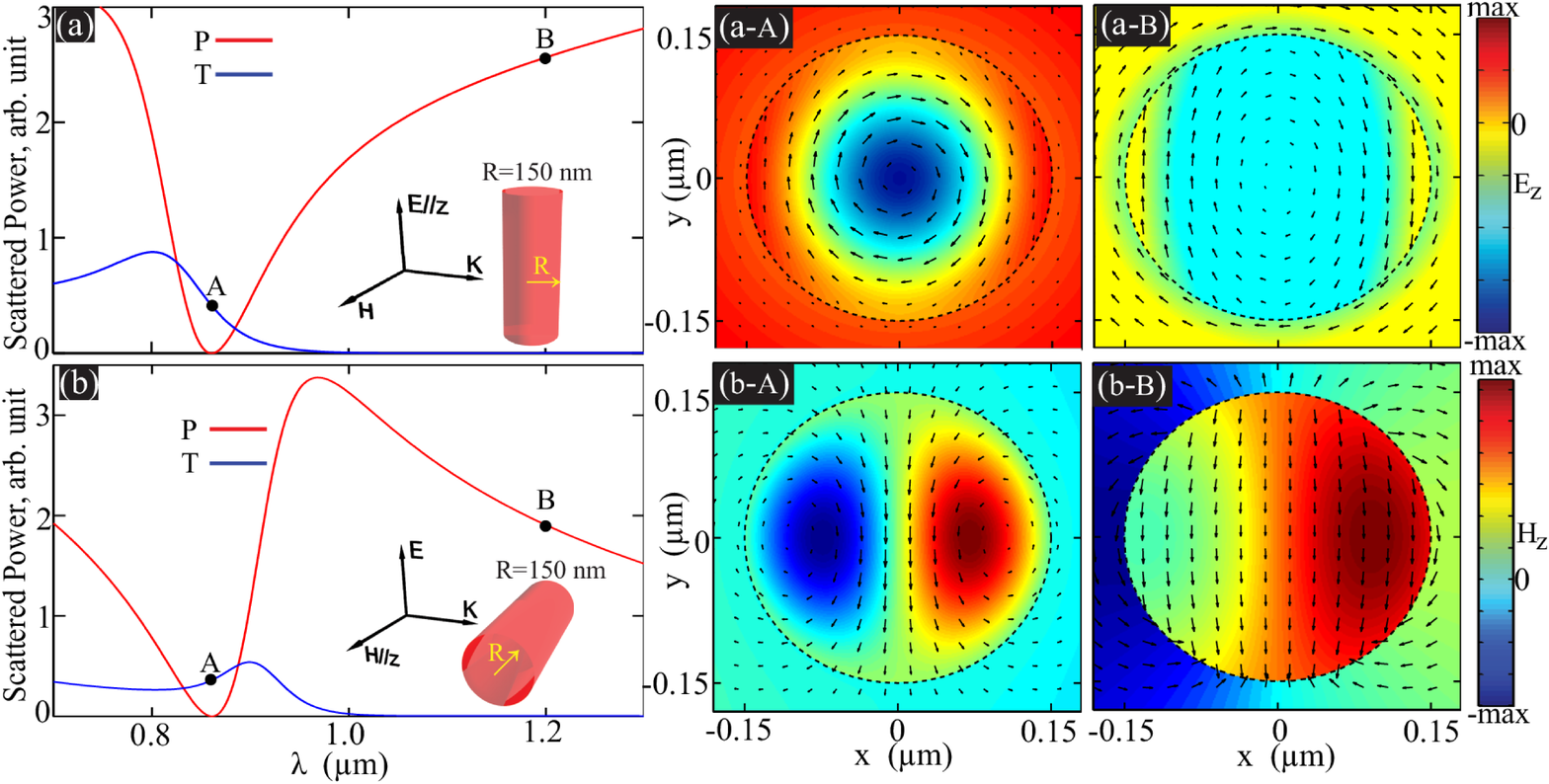}}\caption{\small (Color online) Scattered power spectra for a dielectric ($n=3.5$) nanowire of radius $R=150$~nm for (a) $s$-polarized and (b)  $p$-polarized plane waves. The scattering configurations are shown as insets in (a) and (b) respectively. For both polarizations, the contributions from current-integrated [based on  Eq.(\ref{ED_TD})] ED ($\textbf{P}$, red curves) and TD ($\textbf{T}$, blue curves) are shown. Two points A ($\lambda=860$~nm) and B ($\lambda=1200$~nm) are marked in (a) and (b), and the corresponding near fields are shown in (a-A)-(a-B) for $s$-polarized incident wave [the color-plots and vector-plots correspond to longitudinal electric field ($E_z$) and transverse magnetic field respectively] and in  (b-A)-(b-B) for $p$-polarized incident wave [the color-plots and vector-plots correspond to longitudinal magnetic field ($H_z$) and transverse electric field respectively].}
\label{fig2}
\end{figure*}
%-------------------------------------------------------------------------------

In currently flourishing fields of plasmonics and metamaterials, the scattering by \textit{individual} particles plays a foundational role~\cite{Bohren1983_book,Liu2012_ACSNANO,Liu2014_CPB}.  Nevertheless most demonstrations of the indispensable contributions of TMs have been conducted in engineered structures consisting of relatively complex nanoparticle dimers or clusters~\cite{dong2012toroidal,Huang2012_OE,Ogut2012_NL,Fedotov2013_SR,Fan2013_lowloss,dong2013all,savinov2014toroidal,Basharin2014_arXiv}. Only recently such investigations have been extended to individual nanodisks and nanospheres, which reveal that TMs can not only be effectively excited but also are essential for accurate interpretations for some fundamental scattering properties~\cite{miroshnichenko2014seeing,Liu2014_arxiv}. Similar to the nanodisks and nanospheres, nanowires are also the building blocks for many fundamental structures for various applications~\cite{cui2001nanowire,law2005nanowire,yan2009nanowire}. To introduce TMs into the scattering nanowires may give novel and complete explanations for some exotic nanowire scattering properties, and thus further inspire extra nanowire based applications and offer new platforms for studies of strong light-matter interactions. Such investigations into individual nanowires are vitally important, and have not been discussed so far.
%The incorporation of TMs into nanowire scattering and the following investigations can possibly \frac{expand}{} significantly nanowire based applications and offer new platforms for strong light-matter interaction studies. Such investigations into nanowires are vitally important but currently not available.

In this letter we study the scattering by nanowires based on a Cartesian multipole expansion method with TMs incorporated.  It is found that toroidal dipoles (TDs) can be effectively excited in homogenous dielectric nanowires for both \textit{p}-polarized and \textit{s}-polarized incident waves in the optical spectrum regime. We further show that in core-shell plasmonic nanowires, the  electric dipoles (EDs) and TDs excited can interfere destructively with each other in the far-field, thus rendering the nanowires invisible. It is expected that our new TM-based interpretations of the nanowire scattering could possibly incubate many extra applications based on nanowires and inspire the establishment of new platforms for strong light-matter interactions.

In Fig.~\ref{fig1} we show schematically the structure under consideration. The coordinate system on top shows the configuration of the incident plane wave: it is \textit{s}-polarized ($E$ field along nanowire axis $z$) and \textit{p}-polarized ($E$ field normal to the nanowire axis $z$) for the nanowires in Fig.~\ref{fig1}(a) and Fig.~\ref{fig1}(b) respectively, both of which are assumed to be infinitely long. According to the complete multipole expansion method that has included TMs, the EDs and TDs excited within the nanowires can be expressed by the integrations of the near-field currents [we use the notion of the $\rm exp(-i\omega t)$ for electromagnetic waves]~\cite{Radescu2002_PRE,Fedotov2013_SR}:
%--------------------------------------------------------------
\begin{equation}
\label{ED_TD}
\textbf{P}= {1 \over { - i\omega }}\int d^2r\textbf{j}, ~~\textbf{T}= {1 \over {10c}}\int d^2r[(\textbf{r} \cdot \textbf{j})\textbf{r} - 2r^2 \textbf{j}],
\end{equation}
%-----------------------------------------------------------
where the displacement current  \textbf{j} is:
 %--------------------------------------------------------------
\begin{equation}
\label{current}
\textbf{j}= -i\omega \epsilon_0 [\epsilon(\textbf{r})-1]\textbf{E}(\textbf{r}),
\end{equation}
%-----------------------------------------------------------
 $\epsilon_0$ is the vacuum permittivity; $\epsilon(\textbf{r})$ is the relative permittivity of the nanowire; $\omega$ is the angular frequency of the incident wave; and $c$ is the speed of light.

In Fig.~\ref{fig1}(a) and Fig.~\ref{fig1}(b) we also show  schematically the magnetic field $\textbf{H}$ and current $\textbf{j}$ [or electric field $\textbf{E}$ according to Eq.(\ref{current})] distributions of the TDs excited within the nanowires for both polarizations. As is shown, the TD can be simply interpreted as a series of magnetic dipoles (enclosed circulating currents) aligned along an enclosed path, which actually also represents the field-lines of the induced magnetic field \textbf{H}. The scattering power of ED and TD in Eq. (\ref{ED_TD}) on the $x-y$ plane are respectively: ${{k^3 c} \over {8\epsilon_{0}}}\textbf{P}^2$ and ${{k^5 c} \over {8\epsilon_0 }}\textbf{T}^2$ for \textit{s}-polarization; ${{k^3 c} \over {16\epsilon_0}}\textbf{P}^2$  and   ${{k^5 c} \over {16\epsilon_0}}\textbf{T}^2$ for \textit{p}-polarization. The scattered power spectra of the EDs and TDs excited [calculated according to Eq.(\ref{ED_TD}) within homogenous dielectric (refractive index $n=3.5$) nanowire of radius $R=150$~nm] with incident waves of both polarizations are shown in Fig.~\ref{fig2}(a) and Fig.~\ref{fig2}(b).  It is clear that the TDs can be effectively excited  for both $s$ and $p$ polarized waves [see the blue curves in Fig.~\ref{fig2}(a) and Fig.~\ref{fig2}(b) respectively]. To further verify this, we have selected two points A and B in the scattering power spectra ($\lambda=860$~nm and $\lambda=1200$~nm respectively) and show the near-field distributions at those points in Fig.~\ref{fig2}(a-A)-Fig.~\ref{fig2}(b-B). According to Fig.~\ref{fig2}(a) and Fig.~\ref{fig2}(b), at point A the excitation of ED is negligible and thus there is only TD excitation (modes of magnetic type and of higher orders have not been considered here~\cite{Vynck2009_PRL,kallos2012resonance,Liu2013_OL2621,Liu2014_CPB}). For \textit{s}-polarization the color-plots and vector-plots correspond to longitudinal electric field ($E_z$) and transverse magnetic field respectively [see Fig.~\ref{fig2}(a-A) and Fig.~\ref{fig2}(a-B)]; while for \textit{p}-polarization the color-plots and vector-plots correspond to longitudinal magnetic field ($H_z$) and transverse electric field respectively [see Fig.~\ref{fig2}(b-A) and Fig.~\ref{fig2}(b-B)].

%-------------------------------------------------------------------------------
\begin{figure*}
\centerline{\includegraphics[width=16.5cm]{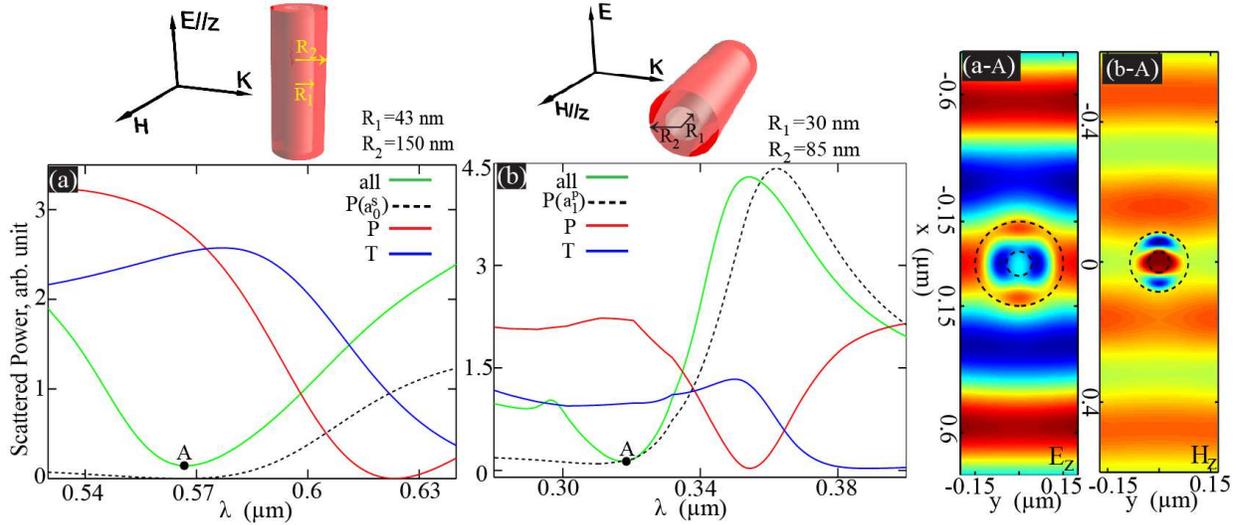}}\caption{(\small Color online) Scattered power spectra for silver core-dielectric ($n=3.5$) shell nanowires of inner radius $R_1$ and outer radius $R_2$ are shown in (a) $R_1=43$~nm and $R_2=150$~nm of $s$-polarization and (b) $R_1=30$~nm and $R_2=85$~nm of $p$-polarization [The curves in (b) are not smooth due to the uneven permittivity data of Ag]. The specific scattering configurations are also included above (a) and (b) respectively. For both polarizations the total scattered power [all, green curves, see Eq.(\ref{C_ext})], and the contributions from far-field deduced ED in Eq. (\ref{conventioanl_dipole}) [black curves, \textbf{P}($a_0^s$) for $s$-polarization and \textbf{P}($a_1^p$) for $p$-polarization], current-integrated ED ($\textbf{P}$, red curves) and TD ($\textbf{T}$, blue curves) are shown. For both polarizations the transparency points A have been marked [$\lambda=566$~nm in (a) and $\lambda=319$~nm in (b)] and the corresponding near-field distributions are shown in (a-A) in terms of $E_z$ and (b-A) in terms of $H_z$ respectively. The wave is propagating from negative $x$ to positive $x$ direction.}
\label{fig3}
\end{figure*}
%-------------------------------------------------------------------------------

For \textit{s}-polarization: (1) At point A, according to Fig.~\ref{fig2}(a-A), within the nanowire region $E_z$ [and thus $\textbf{j}$  according to Eq.(\ref{current})] changes the direction (there are both positive and negative values) and the magnetic field has been mostly confined to the nanowire. This exactly corresponds to current-field distribution for the TD shown in Fig.~\ref{fig1}(a); (2) At point B however, since there is only ED excitation, neither $E_z$ changes its sign nor the magnetic field has been confined within the nanowire [see Fig.~\ref{fig2}(a-B)], which corresponds to a typical ED field distribution.   For \textit{p}-polarization:  (1) At point A, clearly there are both clockwise and anti-clockwise current flows and the magnetic field has been mostly confined within the nanowire [see Fig.~\ref{fig2}(b-A)], which corresponds to a typical current-field distribution of the TD [can be compared to Fig.~\ref{fig1}(b)]~\cite{miroshnichenko2014seeing,Liu2014_arxiv}; (2) In contrast at point B, the field distribution is clearly of a ED type [see Fig.~\ref{fig2}(b-B)].  We note that in sharp contrast to the demonstration of TD excitation in nanowire clusters for only $s$-polarization~\cite{Basharin2014_arXiv}, our work show in even individual nanowires, TD can be effectively excited for both polarizations where the cluster configuration is not required.

So far we have studied ED and TD obtained through current integration and the scattering by nanowires (single or multi-layered) can be solved analytically and the total scattered power of all multipoles is~\cite{Bohren1983_book}:
%%--------------------------------------------------------------
\begin{equation}
\label{C_ext}
%C_{\rm sca}^{\rm s,p} = {4\over {k}}[{(a_0^{\rm s,p})^2} + 2\sum\nolimits_{m = 1}^\infty  {{{(a_m^{\rm s,p})}^2}}],
W_{\rm sca}^{\rm s,p} = 2\epsilon_0\omega^{-1}|{E_0 } c |^2 [{(a_0^{\rm s,p})^2} + 2\sum\nolimits_{m = 1}^\infty  {{{(a_m^{\rm s,p})}^2}}],
\end{equation}
%%-------------------------------------------------------------
where $k=2\pi/\lambda$ is the angular wave-number in the background material (vacuum in this letter); $a_0$ and $a_m$ are the spherical scattering coefficients, where the superscripts $s$ and $p$ corresponds to $s$-polarized and $p$-polarized incident waves respectively. Moreover the scattered fields of the nanowires can be also expressed analytically in terms of $a_0$ and $a_m$. When compared to the far-field scattering of standard electric and magnetic multipoles, the EDs excited with the nanowires for \textit{s}-polarization and \textit{p}-polarization can be deduced respectively as~\cite{Vynck2009_PRL,kallos2012resonance,Liu2013_OL2621}:
%--------------------------------------------------------------
\begin{equation}
\label{conventioanl_dipole}
\textbf{P}(a_0^s)= \epsilon_0 {{4a_0^s} \over {ik^2 }}\textbf{E}_0, ~~~ \textbf{P}(a_1^p)= \epsilon_0 {{8a_1^p} \over {ik^2 }}\textbf{E}_0,
\end{equation}
%-------------------------------------------------------------
where $\textbf{E}_0$ is the electric field of the incident wave. Thus the scattered power from $\textbf{P}(a_0^s)$ and $\textbf{P}(a_1^p)$ are $2\omega^{-1}\epsilon_0\left| a_0^{\rm s} {E_0 } c \right|^2$ and $4\omega^{-1}\epsilon_0\left| a_1^{\rm p} {E_0 } c \right|^2$ respectively. We note here that in the far-field, the ED and TD are indistinguishable since they have identical scattering patterns~\cite{Dubovik1990_PR,Radescu2002_PRE}. The above deduced EDs [see Eq. (\ref{conventioanl_dipole}), which correspond to the overall dipolar scattering] have actually combined the contributions from both the EDs (\textbf{P}) and TDs (\textbf{T}) [see Eq. (\ref{ED_TD})], and some other higher order (in terms of $k$) terms  (such as the mean-square radius of TD) obtained through current integration~\cite{Fedotov2013_SR,Basharin2014_arXiv,miroshnichenko2014seeing,Liu2014_arxiv}. As a result, when ED and TD are tuned to appropriate magnitudes and thus can cancel the scattering of each other,  invisible scattering bodies can be obtained~\cite{Fedotov2013_SR,Basharin2014_arXiv,miroshnichenko2014seeing,Liu2014_arxiv}.

%~\cite{ruan2010temporal,fan2012invisible,mirzaei2013cloaking}

To demonstrate the invisibility induced by the interferences of EDs and TDs excited within nanowires, we focus on the widely studied core-shell plasmonic nanowires. The analytical results are shown in Fig.~\ref{fig3}. The structures under consideration are the silver core-dielectric ($n=3.5$) shell nanowires of inner layer radius $R_1$ and outer layer radius $R_2$ [see the scattering configurations above Fig.~\ref{fig3}(a) and Fig.~\ref{fig3}(b)]. For the permittivity of silver we adopt the experimental data from Ref.~\cite{Johnson1972_PRB}. For \textit{s}-polarization, $R_1=43$~nm and $R_2=150$~nm and the scattered power spectra are shown in Fig.~\ref{fig3}(a). We have included the contributions from ED (\textbf{P}) and TD (\textbf{T}) (red and blue curves respectively), the deduced ED [black curve, \textbf{P}($a_0^s$)] and the total scattered power from all multipoles [green curve, see Eq.(\ref{C_ext})]. At point A ($\lambda=566$~nm) ED and TD  scatter almost the same amount of power and between them there is complete destructive interference, thus leading to more or less null dipolar scattering at this point [see the black curve for \textbf{P}($a_0^s$)]. Moreover, at this point all the other multipole excitations (magnetic dipole and quadrupole for example) have been significantly suppressed, and consequently the total scattering at this point is negligible [see the green curve in Fig.~\rpict{fig3}(a)].  For \textit{p}-polarization with $R_1=30$~nm and $R_2=85$~nm, similar effects can be observed in Fig.~\ref{fig3}(b). The difference is that the minimum scattering point A ($\lambda=319$~nm) is not the point ($\lambda\approx340$~nm) where ED and TD scatter equally and between them there is complete destructive inference. This is due to the fact that for the overall dipolar scattering [black curve Fig.~\ref{fig3}(b)], besides the destructive-interferenced scattering from TD and ED, there are also non-negligible contributions from the mean-square radius of the TD term and some other higher order terms. 

To further verify the induced invisibility, we present near-field distributions at the points marked in Fig.~\ref{fig3}(a) and Fig.~\ref{fig3}(b) in Fig.~\ref{fig3}(a-A) ($E_z$, for \textit{s}-polarization) and Fig.~\ref{fig3}(b-A) ($H_z$ for \textit{s}-polarization) respectively. It is clear that the incident waves have experienced almost no perturbations for both polarizations, verifying the invisibility of the nanowires. Moreover, there are still significant field distributions inside the nanowires at those invisible points, which agrees with the results shown in Fig.~\ref{fig3}(a) and Fig.~\ref{fig3}(b) that there are both ED and TD excitation inside. To summarize, we have demonstrated invisible core-shell nanowires originating from interferences of EDs and TDs excited for both polarizations. We note here that though there have been lots of studies on the transparency phenomena of core-shell nanowires (see Refs.~\cite{ruan2010temporal,fan2012invisible,mirzaei2013cloaking} for example), unfortunately none of them have noticed the roles played by TMs.

To conclude, we study the scattering of normally incident plane waves by nanowires through a complete Cartesian multipole expansion method where TMs have been considered. It is shown that for both \textit{p}-polarized and \textit{s}-polarized incident waves, TDs can be effectively excited within homogenous dielectric nanowires. We further demonstrate the TD induced invisibility in core-shell plasmonic nanowires, which simultaneously have significant internal field distribution and negligible scattering, and thus might play a significant role in non-invasive detections and measurements. Here  we confine our studies to TDs and such investigations can certainly be extended to other higher order TMs. The incorporation of TMs into scattering nanowires may offer new insights into the understanding of many nanowire based light matter interactions, and hopefully can inspire lots of applications of nanowires in sensing, nanoantennas, lasers, photovoltaic devices and so on.

 We acknowledge the financial support from the National Natural Science Foundation of China (Grant numbers: $11404403$, $11304389$ and $61205141$), and the Australian Research Council (FT110100037).

%\bibliography{References_scattering_TD}

%==========================
\end{document}